# One Year of DDoS Attacks Against a Cloud Provider: an Overview

Clément Boin*†, Xavier Guillaume*, Gilles Grimaud†, Tristan Groléat* and Michaël Hauspie†
*OVHcloud, Roubaix, France
†Univ. Lille, CNRS, Centrale Lille, UMR 9189 CRIStAL, F-59000 Lille, France

*Abstract*—Distributed denial of service attacks represents one of the most important threats to cloud-providers. Over the years, volumetric DDoS attacks have become increasingly important and complex. Due to the rapid adaptation of attackers to the detection and mitigation methods designed to counter them, the industry needs to constantly monitor and analyse the attacks they face. In this paper, we present an overview of the attacks that were perpetrated against our infrastructure in 2021. Our motivation is to give an insight of the challenge that DDoS attacks still represent within a large European cloud provider

## I. INTRODUCTION

For several years, distributed denial of service attacks have been one of the most important threats of the modern Internet. All Internet actors, as soon as they occupy positions that are either strategic or important to the public, are susceptible to DDoS attacks [1]. As we have seen during the Covid-19 pandemic, services that are essential to our lives today have been compromised by this type of attack. For example, the French insurer AXA was the victim of a DDoS attack in May 2021 on its information system [2]. In July of the same year, the website of the Ukrainian Ministry of Defense announced that it had suffered such an attack [3]. The motivations of the attackers can be multiple. There are the attackers who launch a DDoS attack using an automated attack system that they rent from a third party for no other specific reason that they can. Attackers can also be motivated by political ideology or a desire to arm a competitor — either financially or by altering its brand image. Attackers can also be willing to perform attacks for personal reasons (*e.g.*, disgruntled employee that was layed off) [4], [5].

Due to its position in the European cloud industry, OVHcloud and its customers may be subject to DDoS attacks. Therefore, OVHcloud needs to provide protection against DDoS attacks that is up to date against most recent attacks. Therefore, attacks performed against the infrastructure must be analysed. This paper present an overview of such analysis on the attacks performed in 2021.

This paper is organized as follows: Section II presents the different works that analyze the attacks on a cloud provider infrastructure. In Section III we give an insight on the methodology used to gather information on the attacks so that they can be analysed. Section IV presents the results of the analysis

This work is also supported by CNRS IRCICA USR-3380

performed on last year attacks. Section V, concludes the paper and gives perspective on how the kind of analysis we performed could be used to ellaborate datasets that can be used to develop now DDoS detection and mitigation techniques.

## II. RELATED WORK

Every year, organisations and companies that deal with DDoS attacks — either by being targets of such attacks or by being a provider of protection solution — publish reports on the development of DDoS over the past year.

By studying the following reports [6]–[11], we could get a insight of important characteristics of DDoS attacks, from these organisations point of view. These characteristics are mainly geographical distribution, used attack vectors and time and throughput characteristics.

We can draw the following conclusions. The two leading countries in terms of the number of attacks, whether directed at infrastructure based in these countries or emanating from infrastructure hosted in these countries, are the United States and China. The most observed attack vectors are UDP and TCP, and according to the reports, they follow each other very closely, so they represent most of the attacks. Akamai report states that attacks using more than one attack vector is an increasingly common phenomenon. Over 90% of attacks are less than 4 hours long and most attacks are small, using about 50,000 packets per second to be carried out. Finally, almost all attacks are operated from Botnets.

Although interesting, these reports are from actors whose activities differ from the ones of OVHcloud. Being a mainly european actor, the geographical aspect of the analysis may not yield the same results, should it be conducted on our data. Moreover, the type of service provided is also quite different as one of the main activity of OVHcloud is to provide baremetal environnement as opposed to specific services. The specific services are the responsability of OVHcloud clients. This makes the attack surface different from a SaaS provider and may show different results on attack vectors.

The next section will present the methodology used to gather information needed to produce reports on attacks characteristics. Those reports will then be analysed and we will draw conclusion with regards to the attacks characteristics specific to OVHcloud.

## III. METHODOLOGY

The statistics presented in this paper are taken from the archives of our DDoS attack detection system. As part of the fight against DDoS attacks, we have set up collection systems at various locations in our infrastructure, which provide realtime network traffic statistics to our detection agents. The data collected are mainly ip flows information similar to NetFlow [12]. If certain properties of the NetFlow analyzer match characteristics of a traffic that is suspicious, we generate alerts. These alerts are effective for as long as the traffic is considered suspicious by our heuristics. It is these alerts that we find in our archiving database. An alert is composed of the IP address of the victim, a start date, an end date, the vector used to carry out the attack and the number of packets and bits per second that make up the alert. As indicated in the literature, an attack can be composed of several vectors, in which case there will be several alerts for the same IP address at a similar time interval.

The main attack vectors we have identified are TCP and its different variations (TCP-SYN, TCP-ACK, etc), ICMP, UDP, IP, NTP, DNS, Fragmentation and Chargen. From our point of view, an attack is a set of malicious network traffic to an IP address of one of our users that can use one or more vectors. So, whether an attack is launched by an attacker using a botnet network hypothetically composed of 500 machines or whether an attack is launched by two separate attackers with 250 machines each and who may or may not have been working together, is for us one and the same attack. Therefore, we had to apply additional processing to the data in the database where the alerts are stored to be able to represent an attack as we have just described it.

To construct the attacks, we used the following methodology: each alert is grouped according to the IP address of the victim. For each group, if two alerts have a start and end date that overlap, it means that they are part of the same attack, so they must be merged. Two cases are possible. First, the two alerts use different attack vectors. In this case, the number of packets and bits received must be calculated for the total duration of each attack and then the number of packets and bits received must be divided by the new duration of this attack, which corresponds to the minimum start date and the maximum end date of the two attacks. For example, in 1, there are three alerts that have the same destination IP.

The first one is an alert that lasts five seconds, the BPS and PPS of this alert is ten, the attack vector used is ICMP, the second one also lasts five seconds, for the same number of PPS and BPS but with the UDP vector, the last one lasts ten seconds, still with a number of PPS and BPS equal to ten, it uses the TCP vector. These three alerts form a single attack, a multi-vector attack (ICMP, TCP and UDP) with a total duration of ten seconds, a PPS and BPS equal to twenty as shown in Figure 2.

The second case is when our detection system updates the number of packets and bits seen per second, in which case both alerts will have the same attack vectors. In this case, we need to count the number of packets and bits on the first potential part where the attacks do not overlap, then take the maximum number of packets and bits for the period where the attacks overlap and then the number for the last potential part where the attacks do not overlap.

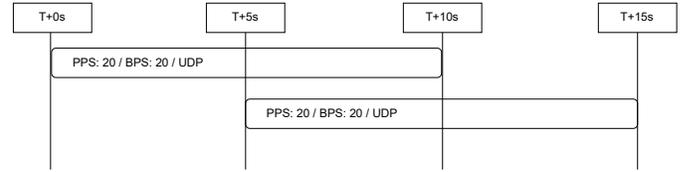

Fig. 1. Alerts with the same attack vector

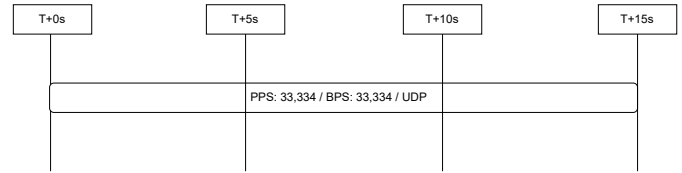

Fig. 2. Result of the merge of the three alerts

## IV. DDoS STATISTICS

In 2021, our DDoS detection systems issued more than three million attacks. Figure 3 represents the number of attacks by month during the year. The month in which the fewer attacks are observed is July, with only a bit over 200,000 attacks, while March and April had over 400,000 attacks each. The median number of DDoS attacks per month observed over the year was just under 264,000 attacks per month. In [8], the authors claim to have observed the same order of magnitude of attacks per month on its DDoS attack protection service. Compared to the figures published by OVHcloud in 2017 on its blog, the most productive month in terms of attacks generated just over 75,000 attacks. An important difference is that in 2017 we accounted for attacks to be detected and mitigated, whereas today we consider an attack as soon as it is detected, it may not be sent to our mitigation systems, for multiple reasons.

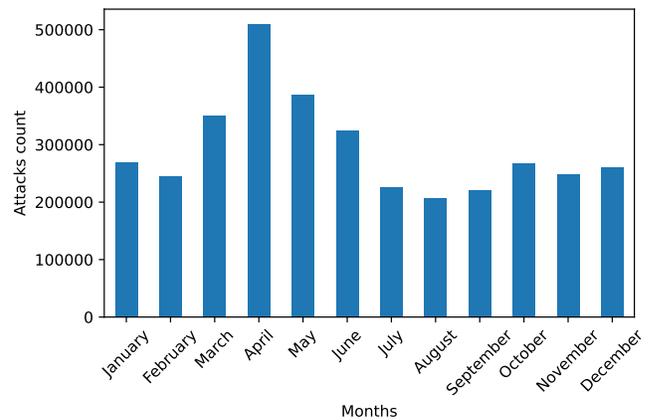

Fig. 3. Monthly observed attacks in 2021

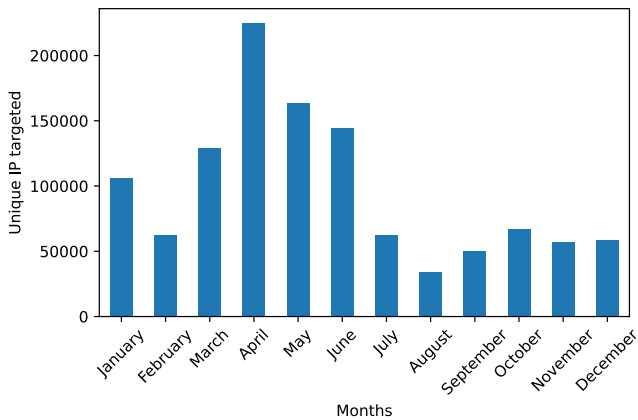 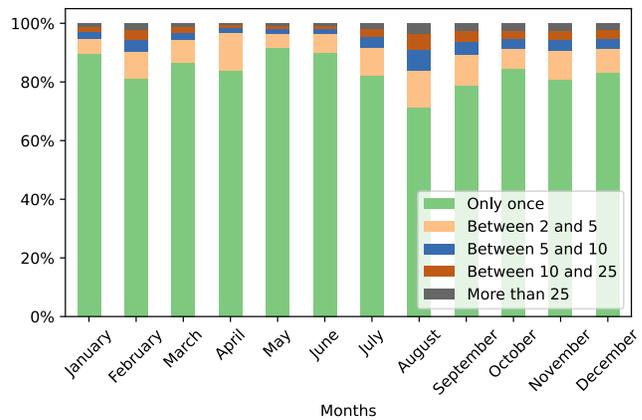

Fig. 4. Unique IP addresses targeted by at least one DDoS in the year 2021

Fig. 5. Frequency of attacks per month and per IP addres

These attacks were directed against over 650,000 of our IPs out of a total of 4 million IPs allocated to our customers. Figure 4 shows the number of distinct IP addresses that have been the victim of at least one attack per month. A slight discrepancy can be observed between Figure 3 and Figure 4, although the trend is the same, and it can be explained by the fact, that some IP addresses are attacked more than one time as shown in Figure 5. On this figure, we can see that for 10% of our customers, DDoS attacks represent a daily threat, as they are targeted several times a month. It is for these customers that our detection and mitigation systems are most important, as DDoS attacks can generally impact their business activities. We can see between the months of March and June, an increase in attack activity is observed.

The month with the highest number of IP address with at least one attack is April with more than 215,000 IPs were attacked while August was the quietest month with 35,000 attacked IPs recorded. The most affected months are March to May. During this period of the year, the public exposure of OVHcloud has strongly increased. This attracted more attackers and explained the net increase in the number of attacks over this period. We also notice that the quietest months in terms of number of attacks correspond to the month where the professional activity is lower because of the holiday periods in Europe. It may be less interesting for the attackers to target services that are little used during these periods.

Figure 6 shows the number of DDoS attacks by time of day. We can see that the attacks are unevenly distributed throughout the day, with a peak of attacks occurring in the early evening between 5pm and 11pm. This distribution implies being able to detect as well as mitigate many attacks in a short period of time. The time of day when the attacks occur is not insignificant, it is generally when people go home and use many online services: video games, streaming platforms, etc. It is during these periods that the impact of an attack on our infrastructure would be more important.

Indeed, the slightest congestion would have a significant impact on the quality of service for all customers. Regarding the attack vectors, as shown in Figure 7 more than 45% of the attacks are syn-flood, as opposed to the figures published by [6] where this attack vector represents 54% of their attacks. TCP-based attacks represent most of the attacks observed at OVHcloud. If we add up the percentages of the different attack vectors known to be used over TCP, we get 55% of the attacks observed while [9] reported more than 75%. UDP represents around 20% of the total attacks, other vectors such as IP or ICMP share the rest.

What is interesting to note is that the most used attack vectors are also those that often rely on IP spoofing, which allows the attacks to stay anonymous. This technique also allows the implementation of amplification and reflection attacks. This technique is widely used by booters, DDoS-as-a-Service providers [6], [9].

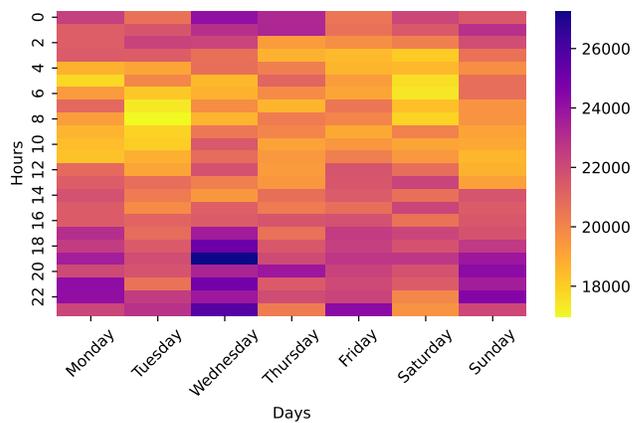

Fig. 6. Occurrence of attacks by day of the week and time (UTC)

By taking a closer look at the IPs targeted by DDoS attacks and the kind of infrastructure, we can draw the following conclusions about the attacked services at OVHcloud. The gaming and e-commerce related infrastructures are most attacked, the same as in our previous analysis in 2017. In fact, if we look at the UDP attack vector that is most widely used for video game protocols, we see a slight increase in the latter during the holiday periods in Europe. Even though games such as Minecraft are mainly based on TCP, the fact that the UDP attack vector increases during certain periods shows that DDoS attacks are often linked to the activity of the target.

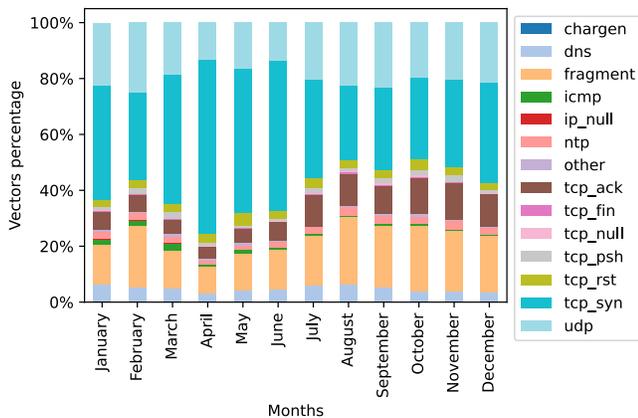

Fig. 7. Attack vectors used in attacks targeting OVH IP addresses in 2021

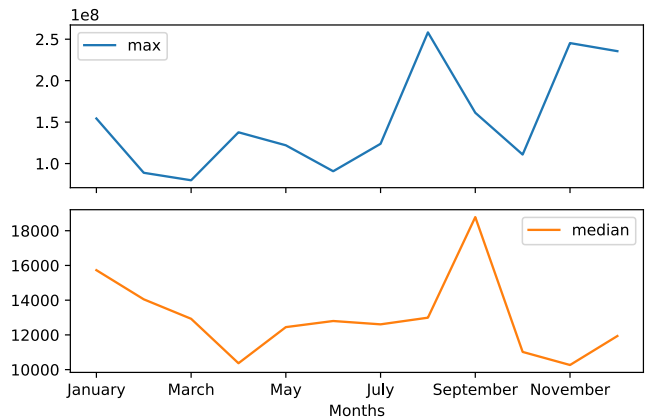

Fig. 8. Average attacks over the months, expressed in packets per second (pps) and the largest attacks per month in terms of pps

The first conclusion is that online gaming services are the most targeted, with Minecraft game servers in the lead. For several years now, we have been observing attacks between game servers, who are engaged in real cyberwar. For example, the Hypixel community server, one of the largest Minecraft server, with over 24M total unique logins to date and a world record 216,000 concurrent players has undergone a campaign of multi-vector attacks with several attacks within this campaign peaking at over 600Gbps [7]. The motivation for attacking game servers seems to be mainly financial. Indeed, the biggest game servers generate revenues, which motivates "administrators" to attack each other to degrade the quality of service at their competitor's, and to convince players to join their servers.

In second place, we find e-commerce platforms. Surprisingly, the targeted platforms are very heterogeneous in size. They range from very large merchants to online shops with moderate traffic. It is rarely a question of rivalry, but rather of extortion. The scheme is quite simple: after an initial attack, the attacker sends an e-mail to his target asking for money him to stop the attacks. For other services, it is more difficult to draw conclusions about the services targeted and the possible motivations of attackers. Startups, public administrations or news websites, everyone seems to be targeted by DDoS attacks, for different reasons: rivalry between competitors, disputes between a service and customers, a desire for censorship in the case of the media, many motives are possible [5].

For the statistics related to packets per second and bandwidth, we can see on figures 8 and 9 that the attacks of the first two months of the year the maximums in terms of pps and bps are correlated. Whereas the months of March and April, which as figure 3 reminds us, are the most active months in terms of number of attacks, are only very representative in the figures. We can therefore deduce that despite the large number of attacks, these attacks were very low in pps and bps. On the other hand, we can see that larger attacks took place during the period from June to September, since although the number of attacks observed during these months is less important, these are the months with the highest number of attacks in pps and bps of the year.

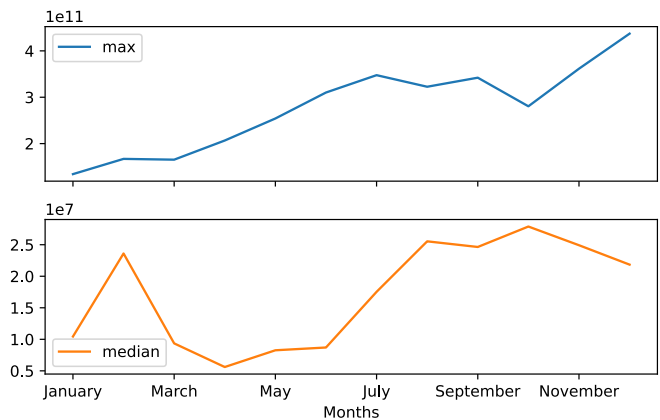

Fig. 9. Average attacks over the months, expressed in bandwidth (bps) and compilation of the largest attacks per month in terms of bps

## V. CONCLUSION

In this paper, we present an overview of the attacks carried against the OVHcloud infrastructure in 2021. This overview can be used to give an example to the scientific community of the challenge that DDoS attacks still represent within a large European cloud provider. The activity of DDoS attacks is globally linked to the specificity of the targets, but also to that of their host. The propagation of attack vectors also seems to be correlated with the sector of activity of the target of these attacks. A service mainly based on HTTP will be targeted by attacks using TCP whereas a video game service is more likely to be attacked via UDP. The figures presented cannot be generalized to all cloud providers as variations in the heuristics used to detect a DDoS attack or even in the business model can for example change from one cloud provider to another. The statistics presented show what it looks like at a cloud provider, and we believe that they may premise the production of datasets that are more in line with what we observe. Therefore, we plan to produce a more accurate comparison between the datasets that are available to the scientific community and the network traffic that we observe in our infrastructure.